\documentclass{JHEP3}

\usepackage{amsmath}
\usepackage{epsfig}

\title{Heat kernel coefficients for compact fuzzy spaces}
\preprint{ YITP-04-64 \\ hep-th/0411029}

\author{Naoki Sasakura \\ Yukawa Institute for Theoretical Physics, Kyoto University,
Kyoto 606-8502, Japan \\ E-mail: \email{sasakura@yukawa.kyoto-u.ac.jp}}

\abstract{I discuss the trace of a heat kernel ${\rm Tr}(e^{-tA})$ for compact fuzzy spaces.
In continuum theory its asymptotic expansion for $t\rightarrow +0$ provides 
geometric quantities, and therefore may be used to extract effective geometric 
quantities for fuzzy spaces. 
For compact fuzzy spaces, however, an asymptotic expansion for $t\rightarrow +0$
is not appropriate because of their finiteness. 
It is shown that effective geometric quantities are found as coefficients of 
an approximate power-law expansion of the trace of a heat kernel 
valid for intermediate values of $t$.
An efficient method to obtain these coefficients is presented and 
applied to some known fuzzy spaces to check its validity.   }

\begin{document}

\renewcommand{\thefootnote}{\fnsymbol{footnote}}
\def\CR{\nonumber \\}
\def\pt{\partial}
\def\be{\begin{equation}}
\def\ee{\end{equation}}
\def\bea{\begin{eqnarray}}
\def\eea{\end{eqnarray}}
\def\eq#1{(\ref{#1})}
\def\la{\langle}
\def\ra{\rangle}
\def\hyp{\hbox{-}}

Fuzzy spaces \cite{Connes, Madore:aq}
have a potential for replacing the classical geometric description of 
space-time. They could incorporate the quantum fluctuation of space-time
which is suggested by the existence of minimal length in semi-classical
quantum gravity and string theory \cite{Garay:1994en,Yoneya:2000bt}
and another kind of arguments \cite{salecker,karolyhazy,Ng:1993jb,Amelino-Camelia:1994vs,
Sasakura:1999xp}.
The scale of the fuzziness is usually identified as the order of Planck length,
and fuzzy spaces are supposed to have the property that classical geometry is 
effective well over the scale.
Since we understand rather well space-time dynamics in terms of geometry, 
it would be useful to understand what is geometry of fuzzy spaces 
in the study of their dynamics.

Since classical geometry can be obtained from trajectories of particles, 
the effective geometry of fuzzy spaces will be determined by low energy propagation modes of 
some fields well over the scale of fuzziness. 
This is in accordance with the spirit of \cite{Chamseddine:1996zu}.
A fuzzy space is often characterized by a scalar field action, which determines the 
propagation of the scalar field. In continuum theory, an asymptotic expansion of 
a heat kernel of a Laplacian has geometric quantities as its coefficients \cite{Vassilevich:2003xt,
Elizalde:1994gf}. For a Laplacian $A=-\Delta$ on a space without boundaries,
the asymptotic expansion for $t\rightarrow +0$ is given by 
\be
\label{asymp}
{\rm Tr}(e^{-tA})\simeq \sum_{j=0} t^{j-\nu/2} \int d^\nu x \ \varepsilon_{2j}(x),
\ee
where $\nu$ is the dimensions of the space, and 
\bea
\label{evalues}
\varepsilon_{0}(x)&=&\frac{\sqrt{g}}{(4\pi)^{\nu/2}},\cr
\varepsilon_{2}(x)&=&\frac{\sqrt{g}}{(4\pi)^{\nu/2}}\frac{R}{6},\\
\varepsilon_{4}(x)&=&\frac{\sqrt{g}}{(4\pi)^{\nu/2}}\left(
\frac{1}{180}R^{abcd}R_{abcd}-\frac{1}{180}R^{ab}R_{ab}+\frac{1}{72}R^2-\frac{1}{30}\nabla_a
\nabla^a R\right),\nonumber
\eea
for $j=0,1,2$.
Therefore, if a Laplacian is given, one can extract some interesting geometric quantities
such as the dimensions of space, its volume, and some curvature integrals through the 
asymptotic expansion \eq{asymp}. 
Especially, $\varepsilon_0(x)$ and $\varepsilon_2(x)$ are important quantities in 
general relativity. When a space has boundaries, there are additional contributions 
to the coefficients from the boundaries, and there can also appear additional 
terms with order of $t$ distinct from \eq{asymp}.

The heat kernel expansion for a non-commutative torus with the Groenewold-Moyal star product
was studied in \cite{Vassilevich:2003yz}. The author has found a power law asymptotic 
expansion for $t\rightarrow +0$, and computed the first four coefficients.
The heat trace coefficients in this case are fully defined by the heat trace expansion 
for ordinary but non-abelian operators. 
The essential difference between the article above and the present paper is that
the main interest of the present one is in compact fuzzy spaces. 
For such a fuzzy space, the number of independent modes is finite, 
and therefore an asymptotic expansion of the trace of a heat kernel for 
$t\rightarrow +0$ is not comparable with \eq{asymp}.
The trace is well-defined on the entire complex plane of $t$, 
including the region $t\leq 0$ in which it diverges in continuum cases.

On the other hand, however, a fuzzy space is expected to approach classical
geometry for low-energy propagation modes. This suggests that there should exist a parameter 
region $t_{min}\lesssim t \lesssim t_{max}$, in which the trace of a heat kernel is in good 
agreement with the expansion \eq{asymp}.  The maximum value $t_{max}$ is related to the
finiteness of the size of a compact fuzzy space. The finiteness will make a Laplacian to 
have discrete spectra, and the trace of a heat kernel will be expanded in
exponentials rather than in powers for sufficiently large $t$.  
The minimum value $t_{min}$ is related to the scale over which the description in
terms of classical geometry is sufficiently good. 
The high spectra of an operator $A$ of a fuzzy space which take significantly different
values from those of a continuum Laplacian
will make large contributions for sufficiently small $t$ enough to 
invalidate the continuum expansion \eq{asymp}.

The aim of the present paper is to show a simple and efficient way to find the parameter region  
$t_{min}\lesssim t\lesssim t_{max}$,
the dimensions and the effective geometric quantities 
from an approximate power-law expansion of the trace of a heat kernel for a compact fuzzy space.
This method will be applied to some known fuzzy spaces, and will be checked numerically.
Presently this method is only suitable for numerical investigations, but could be also used 
in analytical treatment in future study.

To explain the idea, let me define 
\be
\label{defoff}
f(t)=t^{\nu/2}{\rm Tr}(e^{-tA}).
\ee
In continuum theory, this function with $A=-\Delta$ on a space without boundaries
has an asymptotic expansion 
\bea
\label{defofa}
f(t)&\simeq& \sum_{j=0} a_j t^j, \cr
a_j&=&\int d^\nu x \ \varepsilon_{2j}(x),
\eea
for $t\rightarrow +0$. As explained above, since our interest is in intermediate values
of $t$ for a compact fuzzy space, let me take the Taylor expansion of the function $f(s)$ 
about a point $s=t$,
\be
f(s)=\sum_{i=0}^{N-1} \frac{f^{(i)}(t)}{i!}  (s-t)^i +R_N(s,t),  
\ee
where $R_N(s,t)$ is the remainder. This expansion can be rearranged in the form
\be
\label{expoff}
f(s)=\sum_{j=0}^{N-1} a_j^N(t) s^j +R_N(s,t),
\ee
where the `coefficient functions' $a^N_j(t)\ (j=0,\cdots,N-1)$ are defined by
\be
\label{defan}
a^N_{j}(t)=\sum_{i\geq j}^{N-1} \frac{(-t)^{i-j}}{j!\,(i-j)!}\, f^{(i)}(t).
\ee
The expansion \eq{expoff} is comparable with \eq{defofa} with the replacement of the
variable $s$, and $a^N_j(t)$ may be
identified as an approximation of $a_j$ for $j<N$. This approximation is parameterized
by the maximum order of derivative considered and the reference scale parameterized by $t$.

In continuum cases, the limiting values 
$\lim_{N\rightarrow \infty} \lim_{t\rightarrow +0} a^N_{j}(t)$ will provide the 
coefficients $a_j$ of the asymptotic expansion.
In the present context of compact fuzzy spaces, however,  
the expansion \eq{expoff} will be only comparable with
the continuum expansion \eq{asymp} 
in a region $t_{min}\lesssim s,t \lesssim t_{max}$. 
Therefore it is not appropriate to take the limit $t\rightarrow +0$, while there
is also a subtlety in taking $N\rightarrow \infty$, which will be explained in the following
paragraphs.
In fact, what is expected at most is that, with a reasonable choice of $N$, the functions $a^N_j(t)$ take
almost constant values in a range $t_{min}\lesssim t \lesssim t_{max}$, and 
the nearly constant values may be identified as the effective geometric quantities
of a fuzzy space. The validity of this expectation will be explicitly checked in
some examples in due course.

The finite truncation of derivatives parameterized by $N$ is essential for compact fuzzy spaces.
For example, let me take a positive even integer as $\nu$. Then,
since the trace of a heat kernel is regular on the whole complex plane of $t$, 
the values $a^N_j(t)$ approach $\frac{1}{j!}f^{(j)}(0)$ in the limit $N\rightarrow \infty$,
because the Taylor expansion of $f(s)$ about $s=t$ gives 
\be
f^{(j)}(0)=f^{(j)}(t-t)=\sum_{i=0}^\infty \frac{(-t)^{i}}{i!} f^{(i+j)}(t)
=\lim_{N\rightarrow\infty} j!\, a_j^N(t).
\ee
However, these are not the values we want. For example,
$\lim_{N\rightarrow \infty} a^N_0(t)=f(0)=0$, which cannot be identified
as the volume of a compact fuzzy space.

The above fact can be physically explained as follows. 
Because of the regularity of a heat kernel of a compact fuzzy space, the 
remainder $R_N(s,t)$ vanishes in the limit $N\rightarrow \infty$ for a given $s,t$
\footnote{For an odd $\nu$, since $f(s)$ has a singularity at $s=0$, $|s-t| < t$ must
be assumed for $t>0$.}.
This means that the power-law part of \eq{expoff} approximates more precisely the function $f(s)$
by taking higher derivatives into account. But this does not necessarily mean that the power-law 
part approaches the continuum expansion \eq{asymp}.
The power-law part takes also the small scale fuzzy contributions more correctly into account,
and is made to deviate farther from the continuum expansion.
Since higher ultraviolet spectra of $A$ are more subject to the fuzziness,
an effect of this aspect is that the minimum $t_{min}$ is not a constant for a given 
fuzzy space but is shifted to a larger value under taking a larger $N$. 
This will be explicitly checked in an example presented below. 
In another way of saying, it depends on how precisely the spectra of $A$ are observed
whether the classical geometric description is good or not, i.e.,
observing more precisely, one notices more deviation from classical geometry.
Therefore, in the limit of $N\rightarrow\infty$, the parameter region 
$t_{min}\lesssim t \lesssim t_{max}$ will not remain. 
Thus the right way to obtain the effective geometric quantities of a fuzzy space
is to stop at a reasonable order of derivative allowing a certain amount of 
the error $R_N(s,t)$ and fuzziness.
A mathematically more rigorous statement of this aspect deserves to be formulated, 
but presently I have to leave it to future study.

A related matter is the continuum limit of compact fuzzy spaces. 
A compact fuzzy space often has the continuum limit that the dimensions of its 
Hilbert space is taken infinite. 
Accordingly, this continuum limit
must be taken first before the large $N$ limit.    
 
In the following I will present some examples to check the validity and usefulness of the 
method. 

\vskip.2cm
\noindent
(I) Fuzzy spheres \\
A sphere of the unit radius has the following curvature tensor,
\be
R_{abcd}=g_{ac}g_{bd}-g_{ad}g_{bc}.
\ee
Substituting this expression into \eq{evalues}, I obtain
\bea
\label{epsfors3}
\varepsilon_{0}(x)&=&\frac{\sqrt{g}}{(4\pi)^{\nu/2}},\cr
\varepsilon_{2}(x)&=&\frac{\sqrt{g}}{(4\pi)^{\nu/2}}\frac{\nu(\nu-1)}{6},\\
\varepsilon_{4}(x)&=&\frac{\sqrt{g}}{(4\pi)^{\nu/2}}\frac{\nu(\nu-1)(5\nu^2- 7\nu+6)}{360},
\nonumber
\eea
where $\nu$ is the dimensions of the sphere.

A general way of constructing scalar field theory on fuzzy spheres is proposed 
in \cite{Dolan:2003th}.
The authors started with constructing the fuzzy spaces of the orthogonal Grassmannian, 
$SO(\nu+2)/[SO(\nu)\times SO(2)]$.
To select out the modes of a scalar field on a fuzzy sphere embedded in it, 
they constructed an action which 
effectively suppresses all the unwanted modes in the limit of an infinitely large coupling 
constant. The selected modes have the spectra of the second order Casimir of $SO(\nu+1)$,
which are
\be
A_\nu(l)=l(l+\nu-1),
\ee
with the degeneracy,
\be
d_\nu(l)=\frac{(\nu-1+2l)(l+\nu-2)!}{(\nu-1)!\,l!},
\ee
where $l=0,1,\cdots,2L$. The natural number $L$ represents the fuzziness and the continuum limit
is obtained in the limit $L\rightarrow\infty$.

To be specific, let me consider the case of $\nu=3$ (a fuzzy three sphere), 
$L=5$ and $N=5$. In Fig.\,\ref{fig1}, the behavior of $a^5_j(t)\ (j=0,1,2)$ is
shown. There clearly exists a flat region, $0.2\lesssim t \lesssim 0.5$, where $a^5_j(t)$
take almost constant values.
The values of $a^5_j(t)$ in the flat region may be well represented by 
these at $t=0.3$,
\bea
\label{a5value}
a^5_0(0.3)&=&0.443125, \cr
a^5_1(0.3)&=&0.442923, \\
a^5_2(0.3)&=&0.222807, \nonumber
\eea
while the continuum values are
\bea
\label{avalue}
a_0&\approx&0.443113, \cr
a_1&\approx&0.443113, \\
a_2&\approx&0.221557, \nonumber
\eea
which are obtained from \eq{epsfors3} and the volume of a unit sphere
$2 \pi^\frac{\nu + 1}{2}/\Gamma(\frac{\nu+ 1}{2})$. The values of \eq{a5value} and \eq{avalue}
are in good agreement. These consistent values and the existence of 
the flat region support the validity of the present method. 
\FIGURE{
\epsfig{scale=.59,file=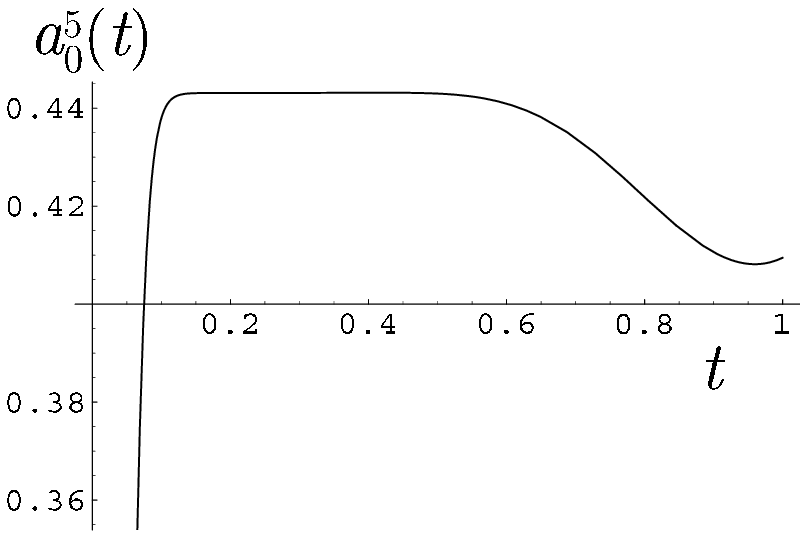}
\epsfig{scale=.59,file=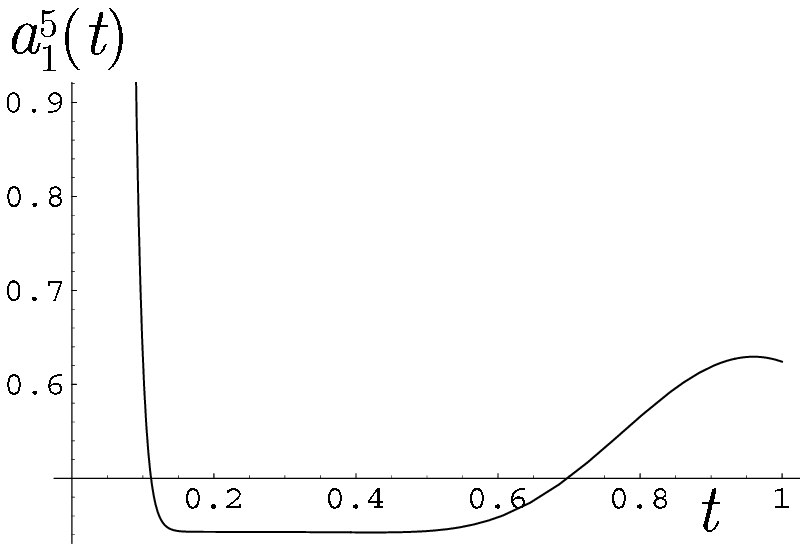}
\epsfig{scale=.59,file=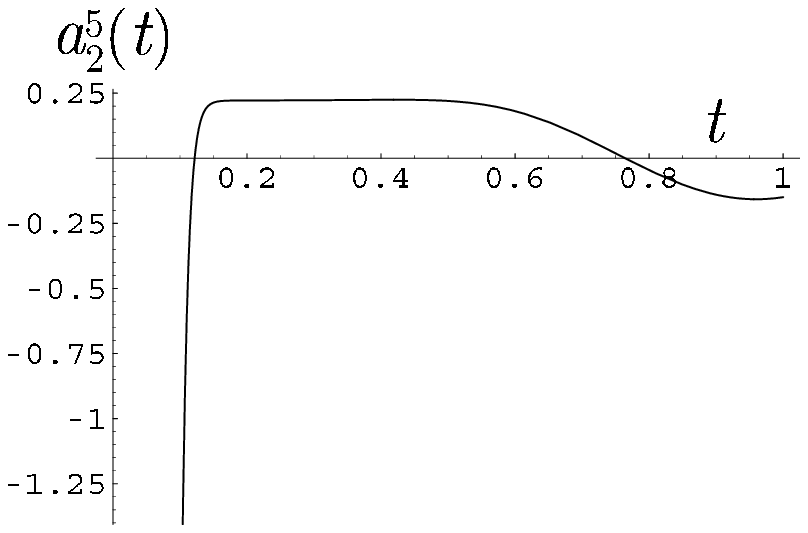}
\caption{The $t$-dependence of the coefficient functions $a^5_j(t)\ (j=0,1,2)$ 
for a fuzzy three sphere.
They take almost constant values at $0.2 \lesssim t \lesssim 0.5$. }
\label{fig1}}

As discussed above, the minimum of the flat region $t_{min}$ is not a 
constant for a given fuzzy space but depends on $N$. This is shown for a 
fuzzy three sphere in Fig.\,\ref{figshift}. 
\FIGURE{\epsfig{scale=.8,file=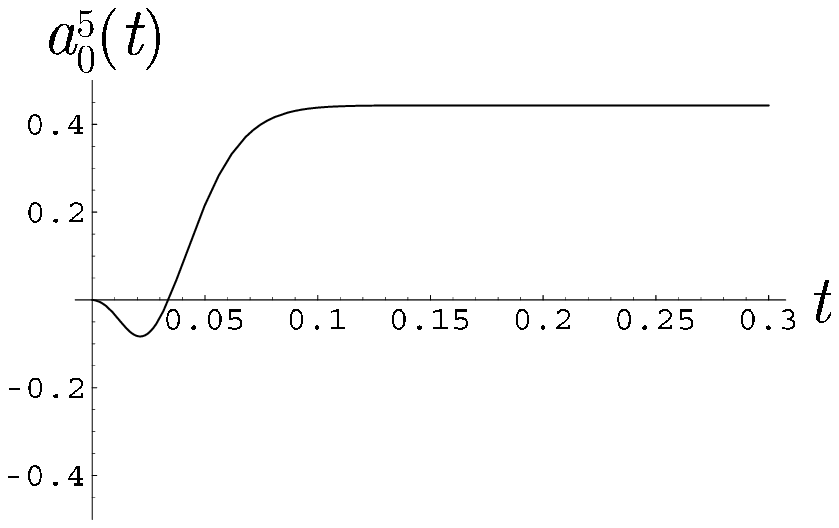} \epsfig{scale=.8,file=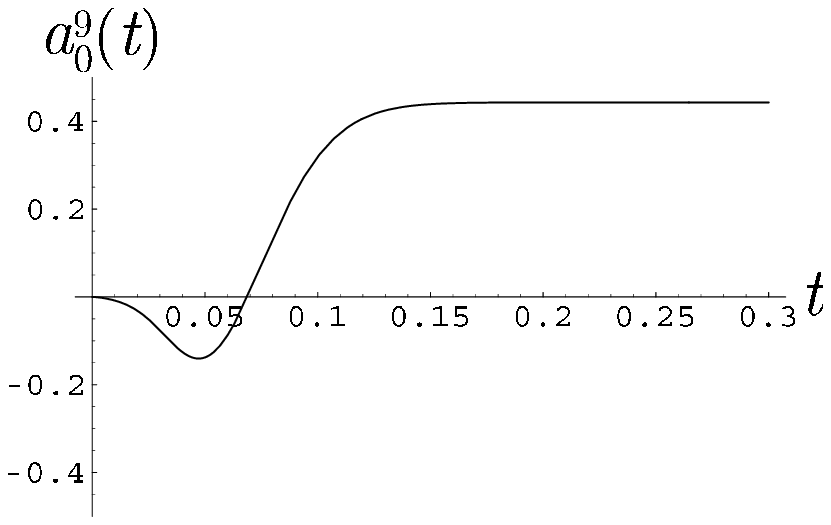}
\caption{The behavior of $a^5_0(t)$ and $a^9_0(t)$ for a fuzzy three sphere with $L=5$. 
The minimum $t_{min}$ is shifted to a larger value under an increase of $N$.  }
\label{figshift}}

\FIGURE{\epsfig{scale=.8,file=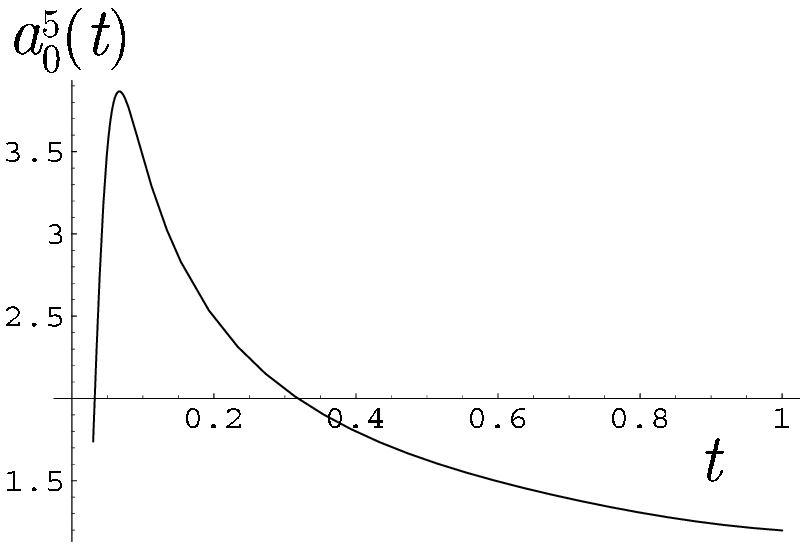}
\caption{The $t$-dependence of the coefficient function $a^5_0(t)$ with a wrong guess
of the dimensions is shown for a fuzzy three sphere. No flat regions can be observed.}
\label{fig2}}
As seen in the definition of $f(t)$ in \eq{defoff}, the dimensions of a space must be known 
before defining the coefficient functions $a^N_j(t)$. In a situation when the dimensions is
not known in the beginning, 
it will be needed to guess the dimensions and try. If the guess is wrong, 
no flat regions will appear, or $a^N_0(t)$ almost vanishes
in a flat region. The former case is shown in Fig.\,\ref{fig2} for a fuzzy three sphere
with a wrong guess of two dimensions, i.e. $\nu=2$ in defining $f(t)$.  
It is clear that the latter case occurs when the guess value 
is larger by an even positive integer.
Therefore the right value of the dimensions can be reached
by checking the existence of a flat region with non-vanishing $a^N_0(t)$.

\vskip.2cm
\noindent
(II) Fuzzy $CP^3$ \\
In \cite{Medina:2002pc,Dolan:2003kq} the authors noted the fact that there are 
different choices of Laplacian on fuzzy $CP^3$. This comes from the two choices of 
the coset space to represent $CP^3$, i.e. $SU(4)/U(3)$ or $SO(5)/[SU(2)\times U(1)]$. 
The spectra are
\be
A(m,n)=2n(n+3)(1-h)+h(n(n+3)+m(m+1))
\ee
with the degeneracy,
\be
d(m,n)=\frac{1}{6}(n-m+1)(n+m+2)(2n+3)(2m+1),
\ee
where $n=0,1,\cdots,L$ and $m=0,1,\cdots,n$. 
The parameter $L$ represents the fuzziness and the continuum limit
is obtained in the limit $L\rightarrow\infty$.
The parameter $h$ interpolates between the $SU(4)$-invariant case at $h=0$ and 
the $SO(5)$-invariant case at $h=1$. I studied the case of $L=20$ and $N=7$ and observed 
a flat region when $\nu=6$, the dimensions of $CP^3$. 
For $0\leq h\leq 1$, the values of the coefficient functions in the flat region
can be well represented by these at $t=0.15$. In Fig.\,\ref{fig3}, $a^{7}_0(0.15)$,
$a^{7}_1(0.15)/(a^{7}_0(0.15))^{2/3}$ and $a^7_2(0.15)/(a^7_0(0.15))^{1/3}$ are plotted
with respect to $h$. The latter two values show no significant changes under the change of $h$. 
Since $a_0$ is proportional to the volume and defines the whole size of a fuzzy space,
the plots suggest that, concerning only the coefficients $a_j$, the change of $h$ can be almost 
absorbed in the rescaling of the metric. 
\FIGURE{
\epsfig{scale=.6,file=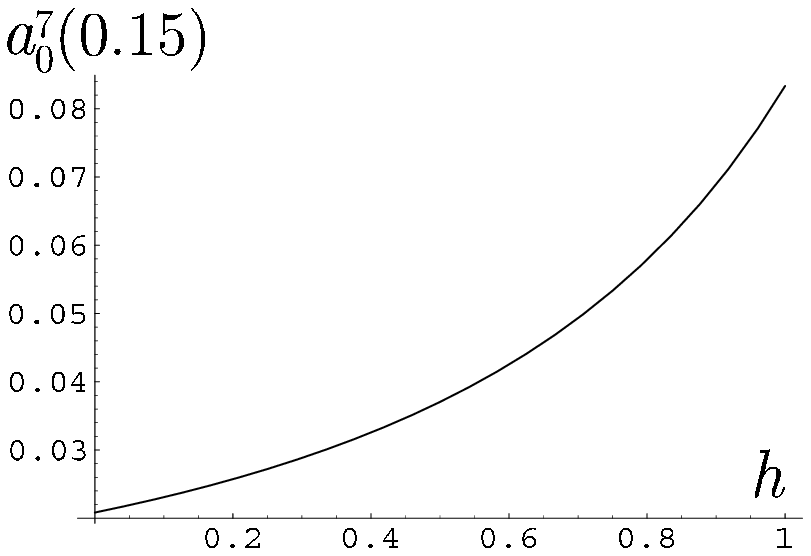}
\epsfig{scale=.6,file=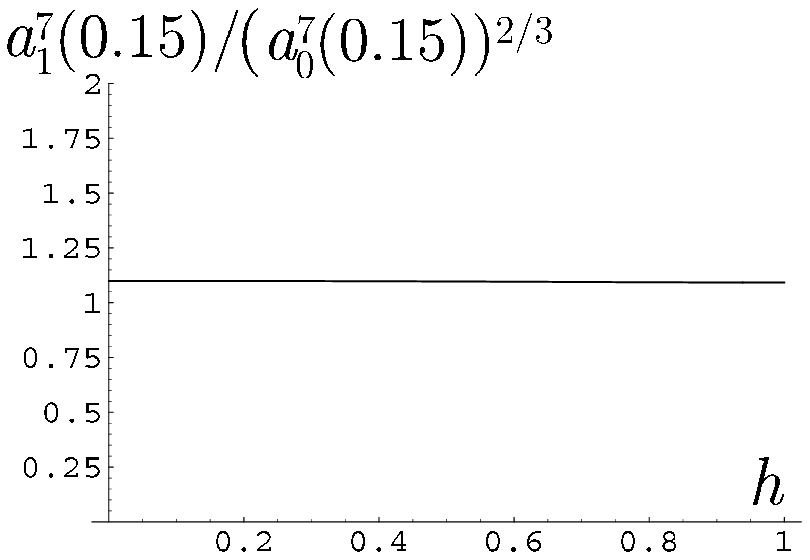}
\epsfig{scale=.6,file=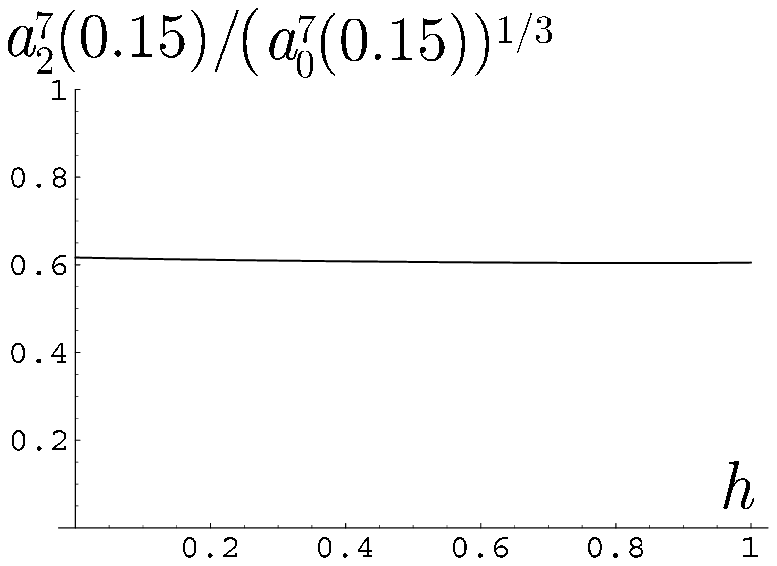}
\caption{The $h$-dependence of $a^{7}_0(0.15)$,
$a^{7}_1(0.15)/(a^{7}_0(0.15))^{2/3}$ and $a^7_2(0.15)/(a^7_0(0.15))^{1/3}$ is plotted for
a fuzzy $CP^3$. The latter two take almost constant values. }
\label{fig3}}

\noindent
(III) Fuzzy $S^1$ and a fuzzy line segment\\
As explained in (I) above, a fuzzy $S^1$ can be obtained from $SO(3)/SO(2)\cong S^2$. 
As same as the fuzzy three sphere shown explicitly in (I), 
the behavior of the coefficient functions $a^N_j(t)$ for a fuzzy $S^1$ 
is in good agreement with the expectation.
Namely, all the $a^N_j(t)$ except $a^N_0(t)$ almost vanish in the flat region. 
There is another reduction of a fuzzy $S^2$ which was studied in another context 
in \cite{Martin:2004dm, Sasakura:2004yr}.
The scalar field action in this case is given by
\be
{\rm Tr}\left( \sum_{i=1}^3 [L_i,\phi]^2 + h [L_3,\phi]^2 \right),
\ee
where $L_i$ are the $SU(2)$ generators. In the limit $h \rightarrow \infty$, 
all the modes with $[L_3,\phi]\neq 0$ are suppressed, and the spectra are effectively given by
\be
A(j)=j(j+1) \ee
with $j=0,1,\cdots,L$ and no degeneracy. In Fig.\,\ref{fig4}, the behavior of the coefficient functions in the case of $L=10$ and $N=5$ 
is shown. 
The value of the dimensions of the space is consistent with $\nu=1$, but 
not only $a^N_0(t)$ but also $a^N_j(t)\ (j>0)$ take non-zero values in the flat region. 
When a space has boundaries, it is known that the coefficients of a heat kernel asymptotic 
expansion have contributions not only from bulk but also from boundaries 
\cite{Vassilevich:2003xt,Elizalde:1994gf}. 
In fact, as was analyzed in \cite{Martin:2004dm, Sasakura:2004yr}, 
the fuzzy space in the limit $h\rightarrow \infty$ is a discretized line segment.
Therefore the non-zero values of $a^N_j(t)\ (j>0)$ in the flat region might be understood
to come from the boundaries of the line segment. 
\FIGURE{
\epsfig{scale=.6,file=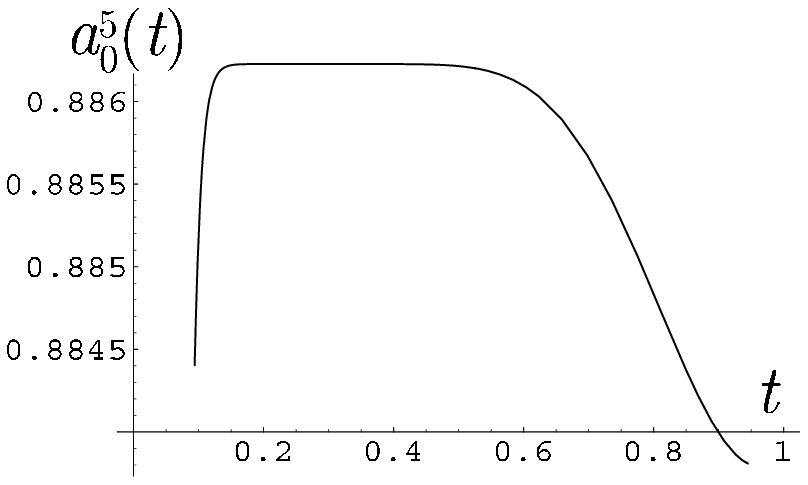}
\epsfig{scale=.6,file=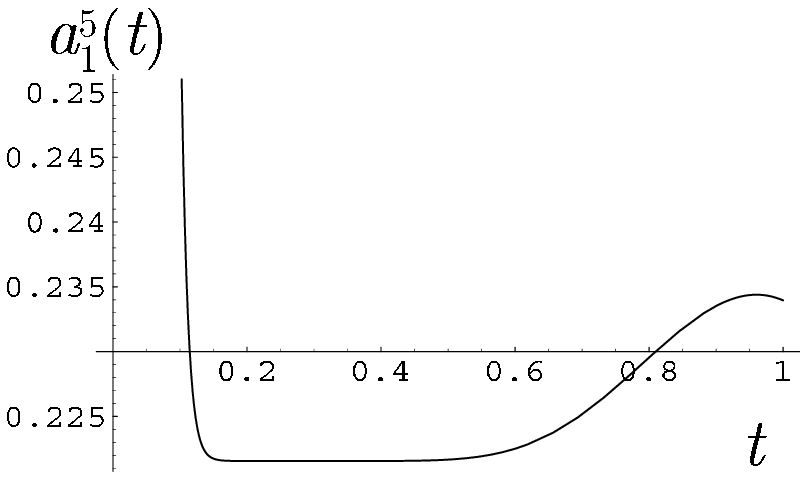}
\epsfig{scale=.6,file=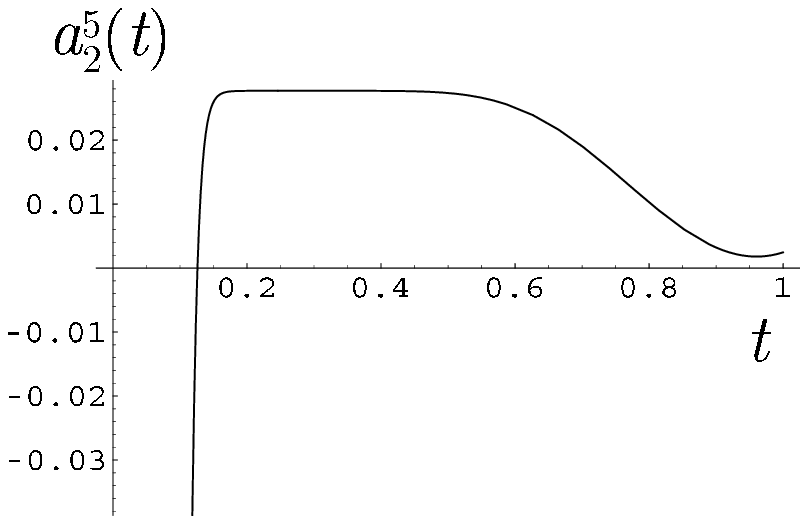}
\caption{The $t$-dependence of the coefficient functions $a^5_j(t)\ (j=0,1,2)$ 
for a fuzzy line segment. There clearly exists a flat region when
$\nu=1$. The functions take non-zero values in the flat region.}
\label{fig4}}

In this paper I have shown that the coefficient functions defined in \eq{defan} can be used
as an efficient method to extract such geometric quantities of a fuzzy space as its dimensions,
volume and curvature integrals. The existence of a flat region in the 
examples shows the classical geometric description is actually effective in a certain range 
of scale. 

A motivation behind of this paper is to understand better the geometric degrees of freedom
of a fuzzy space and its gravitational dynamics. 
In \cite{Sasakura:2004yr,Sasakura:2003ke,Sasakura:2004vm} field theories on evolving fuzzy spaces
are constructed and studied. 
In these papers, however, the evolution of the fuzzy spaces is not dynamical.
I hope that the present work will be helpful to include gravitational dynamics
to these models.


\acknowledgments{The author would like to thank R.\,Sasaki and K.\,Hasebe for explaining $CP^n$,
and the referee for invaluable suggestions.
The author was supported by the Grant-in-Aid for Scientific Research No.13135213 and No.16540244
from the Ministry of Education, Science, Sports and Culture of Japan.}

\end{document}